\documentclass[twocolumn,showpacs,preprintnumbers,amsmath,amssymb]{revtex4}
\usepackage{tabularx,graphicx}

\usepackage{color}
\usepackage{hyperref}
\hypersetup{
    colorlinks=true,
    linkcolor=blue,
    filecolor=blue,      
    urlcolor=blue,
}

\begin{document}
%\documentstyle[aps]{revtex}
%\documentstyle[preprint,aps]{revtex}
%\begin{document}

\newcommand{\beq}{\begin{equation}}
\newcommand{\eeq}{\end{equation}}
\newcommand{\beqn}{\begin{eqnarray}}
\newcommand{\eeqn}{\end{eqnarray}}
\newcommand{\bmath}{\begin{subequations}}
\newcommand{\emath}{\end{subequations}}
\newcommand{\bra}[1]{\langle #1|}
\newcommand{\ket}[1]{|#1\rangle}

%\draft
\title{Momentum of superconducting electrons and the explanation of the Meissner effect}
\author{J. E. Hirsch }
\address{Department of Physics, University of California, San Diego,
La Jolla, CA 92093-0319}

\begin{abstract} 
Momentum   and energy   conservation are  fundamental tenets of physics, that valid
physical theories have to satisfy.  In the reversible transformation between superconducting and normal phases in the presence of a 
magnetic field, the mechanical momentum of the supercurrent has to be transferred to the
body as a whole and vice versa, the kinetic energy of the supercurrent  stays in the electronic degrees of freedom, and no energy is dissipated
 nor entropy is generated  in the process. 
We argue on general grounds that to explain these processes it is necessary that the electromagnetic field mediates the transfer of momentum 
between electrons and the body as a whole, and this requires that when the phase boundary between
normal and superconducting phases is displaced, a flow and counterflow of charge occurs in direction perpendicular
to the phase boundary. This flow and counterflow does not occur according to the conventional
BCS-London theory of superconductivity, therefore we argue that within BCS-London theory
the Meissner transition is a `forbidden transition'.   Furthermore, to explain the
phase transformation in a way that is consistent with the experimental observations requires that (i)
 the wavefunction $and$ charge distribution of superconducting electrons near the phase boundary extend into the normal phase, and (ii) that the charge carriers in the
normal state have hole-like character. The conventional theory of superconductivity does not have
these physical elements, the theory of hole superconductivity does.
 \end{abstract}
\maketitle

\section{introduction}
Experiments \cite{meissner,keesom} and theory \cite{gortercasimir} show that under ideal conditions the superconductor   to normal transition 
 in the presence of a magnetic field is a reversible phase transformation  \cite{reversissue} between equilibrium states of matter
that occurs without energy dissipation and without increase in the entropy of
the universe. In this paper we argue that the conventional BCS-London theory of superconductivity \cite{tinkham,londonbook}
cannot explain how mechanical momentum is conserved in this transition,
and for this reason   BCS-London theory as it stands
is not a viable theory of superconductivity for any superconductor. In other words, BCS theory predicts that the Meissner transition is
a `forbidden transition' \cite{forbidden}, in contradiction with experiment \cite{meissner,keesom}. Instead, we point out that the alternative theory of
hole superconductivity \cite{holesc} explains how the transition occurs in a reversible way conserving mechanical momentum.
The key issue of reversibility and how it is addressed experimentally is discussed  in Appendix B.

We restrict ourselves to non-relativistic electrons, which is sufficient for most solids. In the absence of electric current the average 
mechanical momentum of electrons at any point in space  is zero. 
In the presence of an electric current, the mechanical momentum density of electrons at position $\vec{r}$ is \cite{mstar,freeelmass}
\beq
\vec{\mathcal{P}}(\vec{r})=\frac{m_e}{e}\vec{J}(\vec{r})
\eeq
where $\vec{J}(\vec{r})$ is the current density at position $\vec{r}$, $m_e$ the $bare$ electron mass and $e$ ($<0$) the electron charge.
Consider a  cylinder of radius $R$ and height $h$ in a uniform magnetic field $\vec{H}$ parallel to its axis
pointing in the $\hat{z}$ direction, hanging from a thread of negligible torsion coefficient. 
Assume the material is a type I superconductor with thermodynamic critical field $H_c$  and London penetration depth $\lambda_L$, initially in 
the normal state, and the body is at rest. 
When it is cooled into the superconducting state the magnetic field is expelled from the interior \cite{meissner} (assuming $H<H_c$) 
through the development of a surface current 
\beq
I=\frac{c}{4\pi} hH .
\eeq
 $I$ flows within a London penetration depth of the surface, so the current density is
\beq
\vec{J}=-\frac{c}{4\pi\lambda_L}H \hat{\theta}
\eeq
as follows from Ampere's law $\vec{\nabla}\times\vec{H}=(4\pi/c)\vec{J}$ and the requirement that $\vec{B}=0$ inside the superconductor. Therefore the electrons acquired a non-zero momentum. The momentum density of the supercurrent  is, from Eqs. (1) and (3) 
\beq
 \vec{\mathcal{P}}=-\frac{m_ec}{4\pi\lambda_L e}H \hat{\theta}
\eeq
in a volume $2\pi R \lambda_L h$, hence the total angular momentum of the supercurrent is
\beq
\vec{L}_e=-\frac{m_e c}{2e}hR^2 H \hat{z}  .
\eeq
Note that $L_e$ is proportional to the $bare$ electron mass \cite{freeelmass}.

$L_e$  is a macroscopic angular momentum carried by the supercurrent. For example, for $R=1cm$, $h=5cm$ and $H=200G$, $L_e=2.84 mg \cdot mm^2/s$.
From angular momentum conservation we conclude that the body as a whole must rotate carrying equal and
opposite angular momentum, since the total angular momentum before the system was cooled was zero and no angular momentum was imparted to the system as a whole
upon cooling.
Measurement of the body's angular momentum was never done this way, but instead by applying a magnetic field to an already superconducting body, which will develop a screening current with angular momentum 
equal to those given by Eqs. (3) and (5). Indeed the body is found to rotate with angular momentum given by Eq. (5) 
with opposite sign \cite{kikoin,gyro,doll}. This is called the `gyromagnetic effect'. 

Thus we cannot
doubt that supercurrents carry mechanical momentum and angular momentum. Momentum conservation is a universal law of physics, hence when the state of the system changes so that the supercurrent
changes, the change in the angular momentum of the supercurrent  must be accounted for. 
Quantitatively this momentum is certainly non-negligible.
Consider that a typical current density in superconductors is of order $10^8A/cm^2$, much higher than current densities in normal
metals. It is well known that
for normal state current densities of order $10^6 A/cm^2$ one begins to see significant electromigration effects \cite{electromigr}, where 
the momentum of the conduction electrons is transferred to individual ions causing actual displacement of the ions.
Such effects are not seen in superconductors. Moreover, no Joule heat is dissipated
in the superconductor to normal transition in a magnetic field \cite{keesom,keesom1,keesom2,keesom3}, indicating that there are no
irreversible collision processes that also transfer momentum to the body as in normal conduction. Therefore, 
superconductors need to have a way to transfer the large momentum
of the supercurrent to the body as a whole that is different from the way normal electrons  do it. 
A theory of superconductivity that cannot describe this momentum transfer process cannot account for momentum conservation and hence
cannot be a valid theory
of superconductivity.

          \begin{figure}
 \resizebox{8.5cm}{!}{\includegraphics[width=6cm]{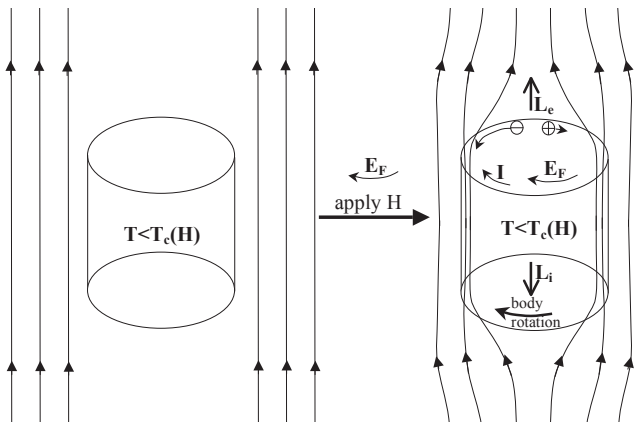}}
 \caption { Process I: a magnetic field is applied to a superconductor at rest. The body acquires angular momentum
 $\vec{L}_i$  antiparallel to the
 applied magnetic field. The supercurrent acquires angular momentum $\vec{L}_e=-\vec{L}_i$ parallel to the magnetic field.
 $E_F$ is the Faraday electric field that exists during the process, which is clockwise as seen from the top, in the same direction
 as the body rotation.
 }
 \label{figure1}
 \end{figure} 
 
            \begin{figure}
 \resizebox{7.5cm}{!}{\includegraphics[width=6cm]{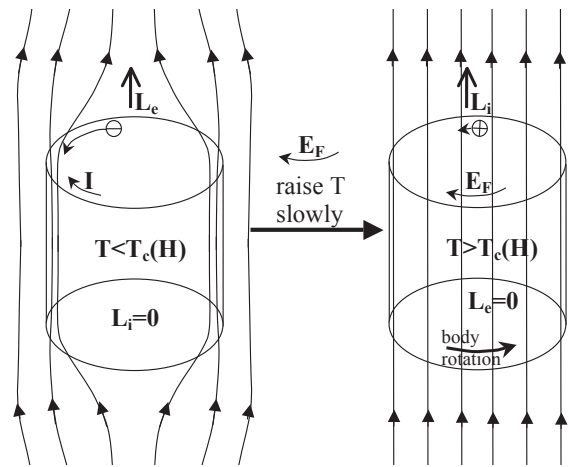}}
 \caption { Process II: a superconductor at rest in a magnetic field turns normal.   The body acquires angular momentum
 $\vec{L}_i$   parallel to the
 applied magnetic field, which equals the   angular momentum $\vec{L}_e$ initially carried by the supercurrent.
 $E_F$ is  clockwise as in Fig. 1, body rotation is counterclockwise.
 }
 \label{figure1}
 \end{figure} 
 
           \begin{figure}
 \resizebox{7.5cm}{!}{\includegraphics[width=6cm]{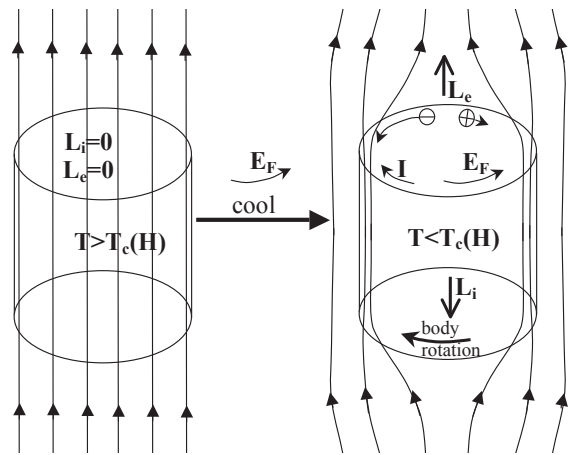}}
 \caption { Process III: a normal metal in the presence of a magnetic field is cooled into the superconducting state.    The body acquires angular momentum
 $\vec{L}_i$   antiparallel to the
 applied magnetic field, and a supercurrent develops with angular momentum  $\vec{L}_e=-\vec{L}_i$ parallel to the
 applied magnetic field.  $E_F$ is  counterclockwise, body rotation is clockwise.
 }
 \label{figure1}
 \end{figure} 
\section{what needs to be explained}
Consider three different processes in which a superconducting cylinder hanging from a thread will acquire angular momentum,
shown in Figs. 1, 2 and 3.
We assume, consistent with the conventional theory of superconductivity and with experiment, that all processes are reversible \cite{reversissue}.

{\it Process I:} The body is at rest at temperature $T<T_c$, and a magnetic field $H<H_c(T)$ is applied (Fig. 1). A clockwise supercurrent develops to prevent the magnetic field from
penetrating its interior, and the body starts to rotate in the  clockwise direction (as seen from the direction where the magnetic field is pointing).

{\it Process II:} The body is at rest  in a magnetic field H and has a clockwise supercurrent preventing the magnetic field from entering its interior. 
Electrons in the supercurrent are moving counterclockwise. The temperature is raised
to slightly above $T_c(H)$, the body enters the normal state, the supercurrent stops and the body starts rotating in counterclockwise direction (Fig. 2).

{\it Process III:} The body is at rest in a uniform magnetic field and initially at temperature $T>T_c(H)$. The temperature is lowered, the body enters the superconducting
state and expels the magnetic field from its interior and starts rotating in clockwise direction (Fig. 3), while electrons in the generated
supercurrent move in counterclockwise direction.

All these processes conserve total mechanical angular momentum (of the electrons in the supercurrent plus the ions in the  body). Since there are no electric fields in the initial and final states, there is no momentum in the electromagnetic field. We also assume that the processes are slow enough that no momentum is carried away by electromagnetic radiation. We will argue that only process I can be explained by the conventional theory of superconductivity. Note that only in process I is the direction of 
the Faraday electric field that develops in the process, $E_F$, parallel to the motion of the ions, in processess II and III it is antiparallel. 
In the following we discuss these three processess.

\subsection{Process I}
As the magnetic field is applied, an azimuthal Faraday electric field develops in the region within $\lambda_L$ of the surface of the cylinder   in clockwise direction, given by
\beq
E_F=\frac{\lambda_L}{c}\frac{\partial H}{\partial t}
\eeq
assuming the magnetic field penetrates a distance $\lambda_L$. 
We have for the velocity of a Bloch electron
\beq
\vec{v}_k=\frac{1}{\hbar}\frac{\partial \epsilon_k}{\partial \vec{k}}
\eeq
where $\epsilon_k$ is the band energy. Within semiclassical transport theory the equation of motion is
\beq
\frac{d\vec{v}_k}{dt}=\frac{1}{\hbar^2}\frac{\partial^2\epsilon_k}{\partial \vec{k}\partial \vec{k}} \frac{d(\hbar \vec{k})}{dt}=  \frac{1}{\hbar^2}\frac{\partial^2\epsilon_k}{\partial \vec{k}\partial \vec{k}} (e\vec{E}_F)
\eeq
Assuming an isotropic band and defining
\beq
\frac{1}{m^*_k} =  \frac{1}{\hbar^2}\frac{\partial^2\epsilon_k} {\partial \vec{k}\partial \vec{k}}
\eeq
the equation of motion is
\beq
\frac{d\vec{v}_k}{dt}=\frac{1}{m^*_k} (e\vec{E}_F)=\frac{1}{m^*_k}  \frac{e\lambda_L}{c}\frac{\partial H}{\partial t}
\eeq
and the change in velocity of the electron when the magnetic field increases from $0$ to $H$ is
\beq
\Delta v_k= \frac{1}{m^*_k}  \frac{e\lambda_L}{c}H
\eeq
The change in electronic momentum is $m_e \Delta v_k$, so the total change in electronic momentum is
\beq
P_e \equiv  \sum_{k\ occ} m_e \Delta v_k = \sum_{k\ occ}  \frac{m_e}{m^*_k} \frac{e\lambda_L}{c}H
\eeq
where the sum over  $k$ here and in what follows is over the  $occupied$ states in the band.
The current density that develops is
\beq
J=\frac{1}{V}\sum_{k\ occ} e\Delta v_k=\frac{1}{V}\sum_{k\ occ} \frac{1}{m^*_k}  \frac{e^2\lambda_L}{c}H
\eeq
and using eq. (3) yields for the penetration depth
\beq
\frac{1}{\lambda_L^2}=\frac{4\pi e^2}{c^2}(\frac{1}{V}\sum_{k\ occ} \frac{1}{m^*_k}).
\eeq
 Using the expression for the current density Eq. (13) to replace the sum over $k$ in Eq. (12)
and the expression for the current density Eq. (3)  yields for Eq. (12)
\beq
P_e=V \frac{m_ec}{4\pi\lambda_L e}H
\eeq
with $V=2\pi R \lambda_L h$, 
in agreement with Eq. (4), and yields Eq. (5) for the total angular momentum acquired by the electrons.

Now the equation of motion for a Bloch  electron is
\beq
m_e \frac{d \vec{v}_k}{dt}=e\vec{E}_F+\vec{F}_{latt}^k 
\eeq
where $\vec{F}_{latt}^k$ is the force exerted by the ionic lattice on the electron of wavevector $k$.  Using Eqs. (8) and (9) we obtain
\beq
\vec{F}_{latt}^k =(\frac{m_e}{m^*_k}-1)e\vec{E}_F
\eeq
and the total force exerted by the lattice on the electrons is
\beq
\vec{F}_{latt} =\sum_{k\ occ} (\frac{m_e}{m^*_k}-1)e\vec{E}_F
\eeq
By Newton's third law, the total force exerted by the electrons on the lattice is then
\beq
\vec{F}_{on-latt} =- \vec{F}_{latt}= -\sum_{k\ occ} (\frac{m_e}{m^*_k}-1)e\vec{E}_F
\eeq
The Faraday electric field also exerts a force on the positive ions. Assuming charge neutrality we have the same number of positive ions (charges) as negative electrons in the band, 
and the total force exerted on the ions (labeled by $i$)  is
\beq
\sum_i \vec{F}_i=\sum_i m_i\frac{d\vec{v}_i}{dt}=\sum_i |e|\vec{E} + \vec{F}_{on-latt} 
\eeq
and using Eq. (19)
\beq
\sum_i \vec{F}_i=\sum_i m_i\frac{d\vec{v}_i}{dt}=  -\sum_{k\ occ} \frac{m_e}{m^*_k}e\vec{E}_F
\eeq
yielding for the total change in ionic momentum
\beq
P_i \equiv \sum_i m_i \Delta v_i=-\sum_{k\ occ} \frac{m_e}{m^*_k} \frac{e\lambda_L}{c}H =  -P_e
\eeq
hence the total angular momentum acquired by the ions is
\beq
L_i=\frac{m_e c}{2|e|}hR^2 H = -L_e .
\eeq

  Thus, for a charge neutral system the electrons and ions acquire equal and opposite momenta and angular momenta, as one would expect.
  The way the ions acquire momentum and angular momentum is partly from the Faraday field itself and partly from the force exerted
  by the electrons on the ions, as seen from Eq. (20). Irrespective of this, the total angular momentum acquired by the body (and the electrons)  is independent
  of $m^*$ and hence of the interactions between electrons and ions, as seen from Eq.  (23).    The interactions between electrons and ions
  only enter in determining the magnitude of the London penetration depth as seen from Eq. (14).
  
  Note that this contradicts the conclusions of Frenkel and Rudnitsky \cite{frenkelrud}, that argued that the fact that gyromagnetic experiments 
  \cite{kikoin,gyro,doll}
  agree with  Eq. (23) with $m_e$ rather than $m^*$   is  by itself  evidence that  electrons carrying the supercurrent are completely `free' from interactions with the lattice.
  Later in this paper we will argue that Frenkel and Rudnitsky's conclusion still was correct, albeit for different reasons.

  \subsection{Process II}

In process II (Fig. 2), the angular momentum of the supercurrent $L_e$ given by Eq. (5) in direction parallel to the applied field has to be transferred in its entirety
to the body as a whole when the system goes normal. In other words, the angular momentum of the electrons has to go from $L_e$ to zero and that of the ions from 
0 to $L_e$.

The angular momentum of electrons and ions will change due to (i) electromagnetic forces, and (ii) interaction between electrons and ions. Let us examine
them in turn:

\subsubsection{ Electromagnetic forces}
As the magnetic field lines enters the body, a Faraday electric field pointing clockwise is generated throughout the interior of the cylinder (Fig. 2),
that tries to prevent the magnetic field from entering (Lenz's law).
This electric field imparts momentum to electrons in counterclockwise direction and to ions in clockwise direction. Thus this momentum transfer is in direction
exactly opposite to what is needed to reach the final state, where the counterclockwise electron current has stopped and the ions rotate
in counterclockwise direction. 

Can there be a magnetic Lorentz force in the azimuthal direction? It can result from $radial$ motion of charge. Since the ionic charge cannot undergo
radial motion, a magnetic Lorentz force cannot be the source of angular momentum for the ions. For the electrons there could in principle be radial
motion, however there is no such motion within the conventional theory of superconductivity.

\subsubsection{ Electron-ion forces}
The Coulomb interaction between electrons and ions can tranfer momentum between the two subsystems. Initially the momentum of the supercurrent is carried by electrons  bound in Cooper pairs. As the system becomes normal, Cooper pairs unbind and become normal quasiparticles, and the supercurrent stops. Within the conventional theory this process
 has been discussed by
Eilenberger \cite{eilen} using the time-dependent  Ginzburg-Landau (TDGL) formalism. A term in the current density describes the current carried by  normal electrons
stemming from the momentum transferred to the normal electron fluid when the superfluid electron density decreases.
Eilenberger  states that  `this momentum then decays with the transport relaxation time $\tau$'. However, such decay would necessarily lead
to Joule heat dissipation and hence irreversibility, therefore this approach cannot be correct \cite{reversissue}. More generally,  any approach that assumes that the momentum of the Cooper
pair is transferred to normal quasiparticles cannot be correct since in normal metallic transport, decay of electric current is necessarily associated with
thermodynamic irreversibility \cite{slaterjoule} and even electromigration for high current densities. 

As already recognized by Keesom \cite{keesom2},  
{\it ``it is essential
that the persistent currents have been annihilated before the material gets
resistance, so that no Joule-heat is developed.''}  The annihilation of the supercurrent has to be accompanied by transfer of the supercurrent momentum to the body as a whole in order to satisfy
momentum conservation, with no energy transfer and no energy dissipation.
In its 60 years of existence, the conventional theory of superconductivity has offered no clue as to how this happens.

One may speculate that transfer of momentum from the supercurrent to the body may occur through phonon emission or scattering by 
impurities. However, these are not $reversible$ processes: in the reverse transformation from normal to superconducting the body
would not be able to transfer its momentum to the supercurrent by reversing the time arrow in these
processes. For further discussion of the reversibility issue, see Appendix B.

\subsection{Process III}

Process III, shown in Fig. 3,  is even more puzzling than process II. Here one has to explain not only how ions acquire momentum opposite to the direction of the force
exerted by the Faraday electric field on ions, but also how electrons acquire  momentum   in direction 
opposite to the direction of the force
exerted by the Faraday electric field on electrons, all without energy dissipation. Within the conventional theory the Eilenberger formalism
can be applied to describe how electrons acquire their momentum, but no mechanism exists for the body to acquire a compensating momentum
in the opposite direction.

In summary,  we have pointed out  that no valid explanation exists in the literature of conventional superconductivity for how momentum is conserved in the processes 
shown in Figs. 2 and 3. 
We argue that any explanation of the momentum transfer between electrons and the body as a whole that involves collisions between electrons and
ions, or impurities, or defects, or phonons, is necessarily a source of irreversibility, which is not observed \cite{keesom}, hence is invalid \cite{revers,disapp}.
The only other way we know to transfer the momentum between the supercurrent and the body is mediated through the
electromagnetic field, as discussed in the next section.

\section{Momentum transfer mediated by the electromagnetic field}
A key aspect of process II   is that the momentum of the supercurrent gets transferred to the body, but its kinetic energy
is not: the kinetic energy of the supercurrent remains in the electronic degrees of freedom, where it is used to pay the price of the
condensation energy in rendering the superconducting electrons normal \cite{londonh,reversissue}.

In most physical interactions, momentum transfer is accompanied by energy transfer. An exception is when magnetic fields are
involved. A charge moving in a magnetic field will change its momentum but not its kinetic energy: the magnetic field doesn't do work
on moving charges since the magnetic Lorentz force $\vec{v}\times\vec{H}$   is perpendicular to the particle's velocity $\vec{v}$. The momentum change of the particle is
compensated by momentum change of the electromagnetic field. The momentum density of the electromagnetic field is given by
\beq
\vec{\mathcal{P}}_{em}(\vec{r})=\frac{1}{4\pi c} \vec{E}\times\vec{H} 
\eeq
with $\vec{E}, \vec{H}$ electric and magnetic fields. 

Thus we  argue  that the process of transfer of momentum of the 
supercurrent to the body without energy transfer must involve the electromagnetic field in an essential way.
It is natural to conclude that the transfer has to happen in two steps: the first step would transfer the momentum of the supercurrent to the
electromagnetic field, and the second step would transfer the momentum from the electromagnetic field to the
body as a whole. Even though both processes may occur concurrently, it is useful to think of  them as
separate processes.

At first sight Eq. (24) does not appear to help, since the electric and magnetic fields at play  in Figs. 2 and 3 are azimuthal and in the 
$z$ direction respectively, resulting in an electromagnetic momentum $\vec{\mathcal{P}}_{em}$  in the radial direction. However the momentum of the
supercurrent and the body are in the azimuthal direction. This then suggests that in processes II and III {\it an electric field in the radial direction exists}. An electric field in the radial direction
and a magnetic field in the $z$ direction will give an azimuthal $\vec{\mathcal{P}}_{em}$.

Consider process II, the annihilation of the supercurrent when the system goes normal. The momentum of the supercurrent, carried by negative electrons, is in counterclockwise direction (Fig. 2).
Assume that in the process of the supercurrent 
disappearing an electric field $\vec{E}$ pointing radially $inward$ is created, thus creating a $counterclockwise$ electromagnetic field momentum
according to Eq. (24). This could be  a ``storage box'' for the momentum of the supercurrent. In a subsequent step,
this momentum of the electromagnetic field would be transferred to the body in a separate process.

           \begin{figure}
 \resizebox{8.5cm}{!}{\includegraphics[width=6cm]{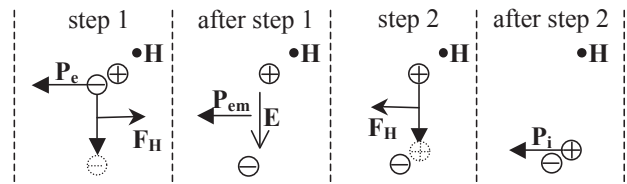}}
 \caption {Illustration of momentum transfer from a negative to a positive charge through the electromagnetic field,
 corresponding to process II, stopping of a supercurrent. Left and down directions correspond to counterclockwise and radially inward directions in Fig. 2.
 Magnetic field $H$ points out of the paper. Initially, the negative charge has momentum $P_e$ pointing to the left, the
 positive charge is at rest. In step 1 the negative charge moves down, the Lorentz force $F_H$ imparts momentum to the right
 cancelling $P_e$. After step 1, the two charges are at rest, and an electric field $E$ exists giving rise to momentum of the
 electromagnetic field $P_{em}=P_e$. The mechanical momentum of the negative charge resides now in the electromagnetic field.
 In step 2 the positive charge moves down and the Lorentz force $F_H$ imparts momentum to it to the left, which is being transferred out of
 $P_{em}$. After step 2 the positive charge carries the momentum $P_i=P_e$ originally carried by the negative charge, and
 the momentum of the electromagnetic field is zero again since $E=0$. 
 }
 \label{figure1}
 \end{figure}

 What we have just described would occur if the process of inward motion of the N-S phase boundary would  also involve
inward motion of negative charge, creating a transitory  inward-pointing electric field, followed by   inward motion of positive charge to retrieve the momentum stored in the field and pass it on to the body.
We show the steps in this process in Fig. 4, where the counterclockwise and inward directions in Fig. 2 corresponds to the leftward 
and    downward directions in Fig. 4 respectively.

In the process shown in Fig. 4, the momentum initially carried by the negative charge is transferred to the positive charge.  There is no assumption on the masses of negative and positive charges, they could be the same or different. If the mass of the positive charge is much larger than that of the negative charge, its speed and
kinetic energy will be much smaller.

In exactly the same (reversed) fashion process III can then be explained, as shown in Fig. 5, namely: if the outward motion
of the phase boundary is associated with $outward$ motion of negative charge, this would create a transitory radially $outgoing$ electric field
and hence a clockwise electromagnetic field momentum that would compensate the counterclockwise mechanical momentum acquired by the
outward-moving electron due to the Lorentz force. In a subsequent step,  positive charge would move outward
and the clockwise momentum of the electromagnetic field would be transferred to the positive charge through the Lorentz force. The end result is negative and
positive charges moving with the same momentum in opposite directions, as shown   in the rightmost panel of Fig. 5. 

           \begin{figure}
 \resizebox{8.5cm}{!}{\includegraphics[width=6cm]{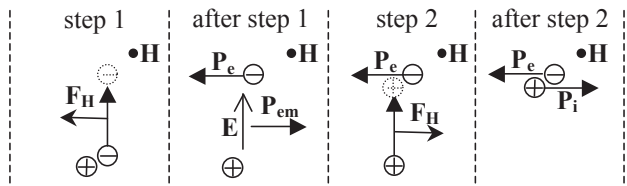}}
 \caption {Similar to Fig. 4 for process III, the Meissner effect. 
 Left and up directions correspond to counterclockwise and radially outward directions in Fig. 3.
 Magnetic field $H$ points out of the paper.
}
 \label{figure1}
 \end{figure}

A problem with these explanations is of course that in a solid, positive ions cannot move radially inward nor outward.
 We will show in the next sections how superconductors get around this problem, through the remarkable properties
 of $holes$. Specifically,  in Figs. 4 and 5  the negative charges correspond to $superconductin$g $electrons$,
 and the positive charges correspond to $normal$ $holes$. We will show that radially moving normal holes transfer 
 azimuthal momentum to the body  without energy dissipation.
 
\section{how supercurrent carriers acquire and lose their momentum without energy dissipation}
           \begin{figure}
 \resizebox{8.5cm}{!}{\includegraphics[width=6cm]{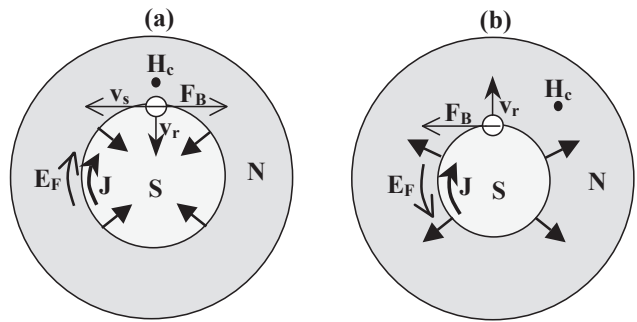}}
 \caption {The normal region is shaded in grey. The magnetic field $H_c$ points out of the paper and
 the Meissner current $J$  flows clockwise, corresponding to counterclockwise motion of electrons. (a) When the phase boundary moves
 inward, the magnetic Lorentz force on a superelectron moving radially inward as it becomes normal is clockwise, thus slowing the superelectron down as it becomes normal. 
 (b) When the phase boundary moves outward, the  magnetic Lorentz force on a normal electron moving radially outward as it becomes superconducting  is counterclockwise, in the
 direction of motion of superelectrons in the Meissner current. In both cases the Faraday field $E_F$ imparts a force in opposite direction.
}
 \label{figure1}
 \end{figure} 
 
The radial motion of negative carriers hypothesized in Figs. 4 and 5 is in the same direction as the
 motion of the phase boundary in the respective situations, as shown in Fig. 6. The momentum parallel to the phase boundary acquired by a negative charge moving a distance $\Delta  x$ in direction
 $\hat{n}$ perpendicular to the phase boundary with radial speed $v_r$ due to the Lorentz force imparted by the magnetic field  is
 \beq
 \Delta \vec{p}_{\parallel}=\int \frac{e}{c}\vec{v}_r\times \vec{H} dt=\frac{e\Delta x}{c}\hat{n}\times{H} .
 \eeq
 We assume that the momentum imparted by  the Faraday electric field (in the opposite direcction) is much smaller and hence can be ignored. We will
 justify this assumption in a later section.
 
 The speed of electrons at the normal-superconductor phase boundary  in an applied magnetic field $H$ is, according to Eq. (11)
 \beq
 v_s=\frac{e\lambda_L}{m_e c} H
 \eeq
 $provided$ we can replace $m^*_k$ by the bare electron mass $m_e$ in Eq. (11). We have recently argued \cite{mstar} that BCS theory itself is inconsistent
 unless the dynamics of electrons in the supercurrent is governed by the bare mass $m_e$ rather than the effective mass, and we will assume
 hereafter that this is the case. Another argument for this will be given in Sect. VI.
 A typical value
 for the superfluid velocity Eq. (26) for $\lambda_L=400\AA$ and $H=500G$ is $v_s=35,225cm/s$.
 
 The momentum of the electron is $m_e v_s$,  hence Eqs. (25) and (26)  indicate that electrons making the transition from normal to superconducting or from superconducting to normal
 advance in the direction of
 the phase boundary motion a distance $\lambda_L$.  For process II, this motion brings  the velocity of the electron in the supercurrent  from $v_s$
 to zero and stores its momentum in the electromagnetic field as shown in Fig. 4 . For process III, this motion gives to the   electron 
 the momentum needed to carry the supercurrent and stores momentum of opposite sign in the electromagnetic field
 as shown in Fig. 5. Thus, this accounts for the transfer of momentum from  electrons to the
 electromagnetic field without energy dissipation, ``step 1'' in Figs. 4 and 5. Next we need to understand how this momentum gets transferred back to the body,
 i.e. the processes denoted ``step 2'' in Figs. 4 and 5, through a `backflow process' which is necessary to preserve local charge neutrality.

 \section{how momentum is transferred   to the body without energy dissipation}
 After step 1 in Figs. 4 and 5  the momentum is stored
 in the electromagnetic field and needs to be retrieved and transferred to the body in step 2. This is achieved through the motion
 of normal  $holes$ in direction perpendicular to the phase boundary.
  
 Consider the two Hall bars shown in Figure 7. They are identical except one has negative and the other positive Hall coefficient $R_H$.
 The Amperian force on the bar is given by
 \beq
 \vec{F}_{Amp}=\frac{I}{c}\vec{L}\times\vec{H}
 \eeq
 where $L=|\vec{L}|$ is the length of the sample and the vector $\vec{L}$ points in the direction of the flow of current  I.
 The Amperian force is of course independent of the sign of the Hall coefficient.
 
             \begin{figure}
 \resizebox{8.5cm}{!}{\includegraphics[width=6cm]{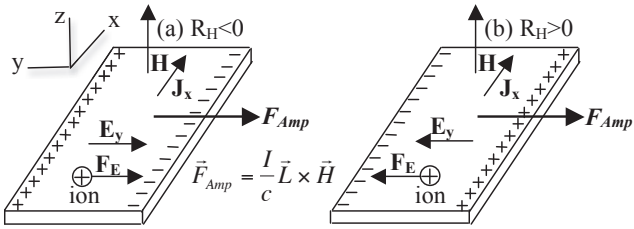}}
 \caption {Hall effect  in metal bars with negative and  positive Hall coefficients $R_H$. The Amperian force on the body is the same independent of the sign
 of the Hall coefficient. However the physical interpretation is very different, as discussed in the text.
}
 \label{figure1}
 \end{figure} 
 For the bar in Fig. 7 (a) the Hall coefficient $R_H$ is negative, the carriers are electrons. The current density is given by
 \beq
 \vec{J}_x= -nev \hat{x}=J_x\hat{x}
 \eeq
 flowing in the positive $\hat{x}$ direction ($e<0$), 
 where $v$ is the magnitude of the drift velocity and $n$ is the density of electron carriers. An electric field pointing to the right 
 (negative $\hat{y}$ direction) exists, given by
 \beq
\vec{E}_y=-\frac{v}{c}H \hat{y}=-E_y\hat{y}
 \eeq
 that equals the magnetic Lorentz force pointing to the left, 
 so that the forces on electrons in the $y$ direction are balanced. Assuming the system is charge neutral, for every conduction electron
 there is a positive charge $|e|$ belonging to an ion that is not moving. The force on this ionic charge is
 \beq
 \vec{F}_{ion}=|e|\vec{E}_y=eE_y\hat{y}
 \eeq
 pointing to the right. The
 total force on all the ions is
 \beq
 \vec{F}_{ion,tot}=nAL |e|\vec{E}_y=nALe\frac{v}{c}H\hat{y}=-\frac{J_xA}{c}LH\hat{y}
 \eeq
 where $A$ is the cross-sectional area of the sample, so the total current is  $I=J_xA$. Hence   in this case
 \beq
 \vec{F}_{ion,tot}=\vec{F}_{Amp}
 \eeq
 For the electrons, electric and magnetic forces are balanced and the electrons move along the $\hat{x}$ direction, hence no
 other force in the $\hat{x}$ direction is acting on electrons. The Amperian force results from the action of the Hall electric field
 $E_y$ on the ions.
 
 The situation is different for the Hall bar with $R_H>0$ shown in Fig. 7 (b).  Here the Hall electric field
 is of opposite sign to the previous case,
  \beq
\vec{E}_y=\frac{v}{c}H \hat{y}=E_y\hat{y}
 \eeq
 pointing in the positive $y$ direction. The current flows in the $x$ direction, hence the net force in the $y$ direction on current carriers
 has to be zero. The force on the ions from the electric field is now
 \beq
 \vec{F}_{ion}=|e|\vec{E}_y=|e|E_y\hat{y}
 \eeq
 pointing {\it to the left}, i.e. in opposite direction to the Amperian force. How does the Amperian force come about?
 
 The answer is, the electrons flowing in the $x$ direction exert a force on the ions, given by
 \beq
 \vec{F}_{e-i}=2eE_y\hat{y}
 \eeq
 so that the total force on the ion is
 \beq
  \vec{F}_{ion,tot}= \vec{F}_{ion}+ \vec{F}_{e-i}=eE_y\hat{y}
  \eeq
  just as in Eq. (30). The total force on the ions is again given by Eq. (32), the Amperian force.
  
  The reason the electrons exert a force on the ions is that the ions exert a force on the electrons that are moving carrying the current.
  According to the semiclassical equations of motion for Bloch electrons the motion of electrons in solids results from the combined
  action of the external force and the force exerted by the ions on the electrons. A detailed analysis is given in Appendix A.
  
This then implies that in a Hall bar with positive Hall coefficient where the drift velocity of current carriers (holes) is $\vec{v}_d$ in the direction of current flow, for each hole
  that moves a distance $d$,
it takes a time interval $\Delta t=d/v_d$ and the momentum transferred in that time from the electrons to the ions is
\beq
\Delta \vec{P}_{ion}=-2\frac{e}{c}\vec{v}_d \times \vec{H} \Delta t=2\frac{e d}{c} \hat{v}_d\times\vec{H} .
\eeq
This momentum transfer from electrons to ions occurs without any irreversible scattering processes, and can only occur when the carriers are holes. 

             \begin{figure}
 \resizebox{8.5cm}{!}{\includegraphics[width=6cm]{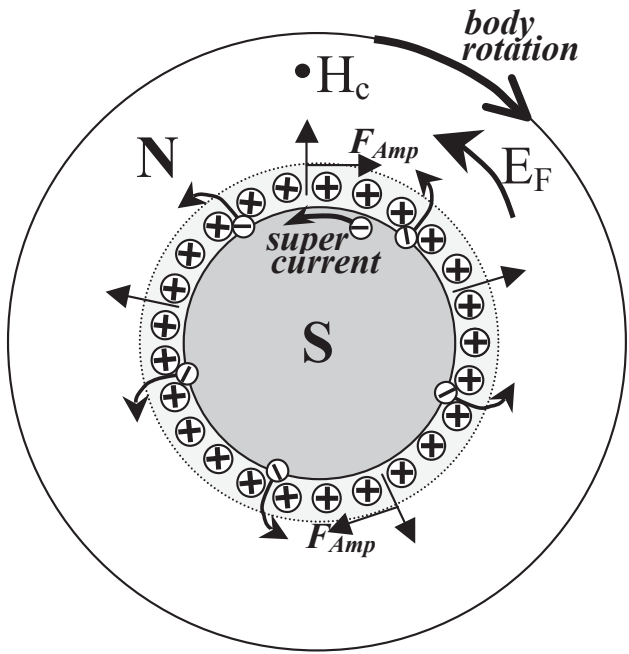}}
 \caption {Expansion of the superconducting phase (grey region) in magnetic field pointing out of the paper. Electrons acquire counterclockwise momentum
 as they thrust outward through the action of the magnetic Lorentz force. A hole current flows outward in a boundary layer of thickness
 $\lambda_L$ (light grey ring). The Amperian force on this current $F_{Amp}$  transfers momentum to the body as a whole in clockwise direction,
 just like for the Hall bar in Fig. 7 (b).}
 \label{figure1}
 \end{figure}

 We can now understand how the momentum transfers between the supercurrent and the body shown in  Figs. 4 and 5 occur in the context of the superconductor to normal transition and
 normal to superconductor transition, i.e. processes II and III. We will describe it for the Meissner effect, process III, Figure 5. In the cylindrical
 geometry the process is shown in Figure 8.  
 
 We assume the superconductor-normal phase boundary is moving radially outward at speed $\dot{r}_0$. An azimuthal electric field pointing counterclockwise
 is generated at and near the phase boundary \cite{dynann}:
 \beq
 \vec{E}_F=\frac{\dot{r}_0}{c}H_c \hat{\theta}
 \eeq

 (1) Step  1: As the phase boundary advances into the normal region, electrons that were in the normal state become superconducting.
 We assume that the electron becoming superconducting   thrusts radially outward (up in Fig. 5) at a high speed $v_r$ a distance $\lambda_L$. In the process it acquires through the action of the magnetic Lorentz force
 an azimuthal  momentum in counterclockwise direction (to the left in Fig. 5)
 \beq
 \Delta \vec{p}_e=-\frac{e\lambda_L}{c}H_c\hat{\theta} +e\frac{\dot{r}_0}{c}H \Delta t \hat{\theta}.
 \eeq
 The second term in Eq. (39) arises from the action of the Faraday field Eq. (38). $\Delta t$ is the time it takes the electron to move a distance $\lambda_L$ at speed $v_r$,
 $\Delta t=\lambda_L/v_r$. Under the condition 
 \beq
 \dot{r}_0<<v_r
 \eeq
the second term in Eq. (39) is much smaller than the first term and we assume it can be neglected, so that Eq. (39) gives the required momentum of the
electron in the supercurrent Eq. (26).
 
   (2) After step 1:  the outward motion of electrons in step 1 creates a radially outward electric field in a boundary layer of thickness $\lambda_L$
   in the normal region, and stores azimuthal momentum in the electromagnetic field, as shown in the second panel in Fig. 5.
   The radial electric field drives a radial outflow of current in this boundary layer of thickness $\lambda_L$ (Fig. 8, light grey ring) that moves at the same
   speed as the phase boundary motion. 
   
   (3) Step 2: We assume that the normal current is carried by
   hole carriers. Normal current flows radially outward in the boundary layer and exerts a force on the body. This force is  the
   Amperian force $\vec{F}_{Amp}$ shown in Fig. 7 (b) and in Fig. 8. It transfers the momentum stored in the electromagnetic field to the body as a whole,
   without energy dissipation, so the body acquires rotational velocity in the clockwise direction.

 (4) After step 2: the momentum acquired by the electron going superconducting in the counterclockwise direction is exactly compensated by the 
 momentum transferred to the body in the clockwise direction.  
 
 The fact that the momenta are exactly compensated does not need proof, it follows from the fact that the backflow propagation of the 
 holes is exactly radial because of the balance of forces: the total azimuthal force acting on the outflowing hole with speed $\dot{r}_0$  is
 the sum of the clockwise magnetic Lorentz force and counterclockwise Faraday force which equals zero:
 \beq
 \vec{F}_{hole}=-|e|\frac{\dot{r}_0}{c}H_c \hat{\theta}+|e|\vec{E}_F=0 ,
 \eeq
 Nevertheless, let us verify that momentum conservation holds. The momentum acquired by an electron going superconducting and
 thrusting radially outward a distance $\lambda_L$ is
 \beq
 \Delta \vec{p}_e=-\frac{e\lambda_L}{c}H_c\hat{\theta}
 \eeq
 neglecting the second term in Eq. (39) under the assumption that $v_r>>\dot{r}_0$.  The `backflow' normal holes
 move at speed $\dot{r}_0$ in the $+\hat{r}$ direction and traverse the boundary layer of thickness $\lambda_L$ in time
 $\Delta t=\lambda_L/\hat{r}_0$. The net force per carrier exerted on the lattice during that time is given by Eq. (36), where
 $E_y$ is the Faraday field $E_F$ and the $\hat{y}$ direction is the $\hat{\theta}$ direction
 \beq
 \vec{F}_{ion,tot}=e\vec{E}_F=e\frac{\dot{r}_0}{c}H_c \hat{\theta}
 \eeq
 which is the same as the Amperian force in Fig. 7 (b). Hence the net momentum transferred to the ions per electron going superconducting is
 \beq
 \Delta\vec{p}_i= \vec{F}_{ion,tot} \Delta t=\frac{e\lambda_L}{c}H_c\hat{\theta}
 \eeq
 equal and opposite to Eq. (42), as expected.

 Exactly the same steps in reverse explain how as the superconducting region shrinks the mechanical momentum of an electron in
 the supercurrent that becomes normal   is transferred to the   body through a radially inward flow of holes in a boundary layer
 of thickness $\lambda_L$.
 
 Returning to the case of the expanding superconducting phase in Fig. 8, we also need to consider the effect
 of the Faraday field in the superconducting region within $\lambda_L$ of the phase boundary, where the supercurrent flows.
 Its effect on the electrons in the supercurrent is to slow them down (force acts in clockwise direction), so that they eventually come
 to a stop when the boundary has moved beyond a distance $\lambda_L$, as discussed in ref. \cite{dynann}. Its effect on the 
 body is to impart momentum in counterclockwise direction, which partially compensates the momentum transfer Eq. (44), 
 resulting in a net transfer of momentum which generates the body's rotation, as discussed in the next section.
  
\section{Macroscopic torque}
Let us now analyze how the macroscopic rotation of the body comes about.
The Amperian force per unit volume exerted on a radially outgoing current $J_r$ in the presence of magnetic field $H$ is
\beq
\vec{F}_{Amp}=-\frac{H}{c}J_r\hat{\theta}
\eeq
where $\hat{\theta}$ is positive in counterclockwise direction. The radial hole current is given by
\beq
\vec{J}_r=n_s|e|\dot{r}_0\hat{r}
\eeq
and   this current occupies  a boundary layer of thickness $\lambda_L$, with volume $V=2\pi r_0\lambda_L h$, with $h$ the height of the cylinder. Hence the torque exerted by the
Amperian force on the boundary layer of thickness $\lambda_L$ flowing outward with speed $\dot{r}_0$ is
\beq
\vec{\tau}_1=\frac{H_c}{c}2\pi r_0^2\lambda_L h n_s e \dot{r}_0 \hat{z}
\eeq
pointing in the $-\hat{z}$ direction, i.e. opposite to the direction of the magnetic field. 

There is also a countertorque due to the clockwise force exerted by the Faraday
electric field $E_F$ on the ions in the superconducting region within distance $\lambda_L$   of the superconductor-normal phase boundary,
where supercurrent flows. The Faraday field in that region is given by \cite{dynann}
\beq
\vec{E}_F(r)=\frac{H_c}{c}\dot{r}_0 e^{(r-r_0)/\lambda_L}\hat{\theta}
\eeq
and it exerts a torque
\beq
\vec{\tau}_2=-2\pi n_s e h \int_0^{r_0}E_F(r)r^2 dr \hat{z}
\eeq
on the body. Doing the integral and assuming $r_0>>\lambda_L$ yields
\beq
\vec{\tau}_2=-\frac{H_c}{c} 2\pi n_s e h \lambda_L[r_0^2-2r_0\lambda_L+2\lambda_L^2]\hat{\theta}
\eeq
so that the net torque on the body  is (neglecting the higher order term proportional to $\lambda_L^3$)
\beq
\vec{\tau}=\vec{\tau}_1+\vec{\tau}_2=\frac{H_c}{c}4\pi \lambda_L^2 h n_s e r_0 \dot{r}_0 \hat{z}.
\eeq
By conservation of momentum, the ionic angular momentum is minus the electronic angular
momentum Eq. (5)
\beq
\vec{L}_i=-\vec{L}_e=\frac{m_ec}{2e}hr_0^2 H_c \hat{z}
\eeq
and the associated torque is
\beq
\vec{\tau}_i=\frac{d \vec{L}_i}{dt}=\frac{m_ec}{e}hr_0 \dot{r}_0 H_c \hat{z} .
\eeq
Equating Eq. (53) to the net torque exerted on the body, Eq. (51), we find
\beq
\frac{1}{\lambda_L^2}=\frac{4\pi n_s e^2}{m_e c^2} .
\eeq
This is the well-known expression for the London penetration depth \cite{tinkham}.
On the other hand, we find from our formula for Bloch electrons Eq. (14) for either a band
close to empty or close to full
\beq
\frac{1}{\lambda_L^2}=\frac{4\pi n_s e^2}{m^* c^2} .
\eeq
where $n_s$ is 
\bmath
\beq
n_s=\frac{1}{V}\sum_{k\  occ}1
\eeq
for a band close to empty, or 
\beq
n_s=\frac{1}{V}\sum_{k\   unocc}1
\eeq
\emath
for a band close to full, and $m^*=m^*_k$ at the bottom of the band for an almost empty band, or
$m^*=-m^*_k$ at the top of the band for an almost full band.

In deriving the expression Eq. (55) for the London penetration depth, we assumed that
superconducting carriers respond to the induced Faraday field as if they were
Bloch electrons with effective mass $m^*_k$, Eq. (9). However, to satisfy momentum conservation 
we found  here that the London penetration depth has to be given by Eq. (54) with the $bare$
electron mass $m_e$. The implication of this is inescapable: our original assumption Eq. (10) leading to Eq. (55)
was incorrect. 
Unlike normal Bloch electrons,
superconducting carriers respond to external field with their bare electron mass, in other words
they are completely `undressed' from the electron-ion interaction. We recently
reached this same conclusion through a completely different path, by examining inconsistencies
within conventional BCS-London theory \cite{mstar}.

In summary,  the macroscopic rotation of the body when the superconducting region expands results from the  torque exerted by the
radially outgoing hole current Eq. (46) on the body  in the clockwise direction exceeding the countertorque
Eq. (50)  exerted by the Faraday electric field on the ions in the 
counterclockwise direction in the region where supercurrent flows by the amount given by
the net torque Eq. (51). For a shrinking superconducting region all the signs are simply reversed.

\section{new physics of superconductivity}

In the previous sections we have described a plausible way to explain the momentum transfer between the electronic degrees of freedom and the body as a whole
in a reversible way in processes II and III. We don't know any other possible way to do this, and no other way has been proposed in the literature. Next let us consider what is
required of a microscopic theory of superconductivity to allow this to occur. We argue that the following are {\it necessary conditions}:

(i) The wavefunction {\it and charge distribution} of superconducting electrons close to the phase boundary extend into the normal state.

(ii) The charge carriers in the normal state that are condensing to give rise to the supercurrent in the superconducting state are $holes$.

Requirement (i) follows from the fact that we assumed in the previous sections that when electrons go from normal to superconducting they  `thrust' into the 
normal region a finite distance $\lambda_L$, thereby acquiring  the momentum needed for the supercurrent through the magnetic Lorentz force. Within BCS theory it is assumed that the superconducting order parameter does leak into the normal region,
leading e.g. to Josephson effects and proximity effects, however BCS theory does not predict that the order parameter has any $charge$ associated with it. This is because charge has to have a $sign$ (negative
or positive), and BCS theory is intrinsically electron-hole symmetric, so the order parameter is not
associated with either negative or positive charge.
Therefore, BCS theory does not satisfy this requirement.

BCS theory also does not satisfy requirement (ii), since within BCS the normal state 
carriers may be electron-like or hole-like. 

Therefore, we conclude that BCS theory does not have the physical elements   required to explain the reversible momentum transfer between electrons and the 
body that takes place in the superconductor-normal and normal-superconductor  transitions in a magnetic field.

Instead, the theory of hole superconductivity \cite{holesc} does have those physical elements, as discussed in earlier papers
\cite{revers,dynann,meissnerexp} and recounted
briefly in what follows:

(i) Within the theory of hole superconductivity electrons in the condensate reside in mesoscopic orbits of radius $2\lambda_L$ \cite{bohr}, while 
they reside in microscopic orbits of radius $k_F^{-1}$ in the normal state. Thus, when electrons go from normal to superconducting they
expand their orbits to radius $2\lambda_L$, and since they are at the normal-superconductor phase boundary this is associated with
negative charge leaking into the normal region. The azimuthal velocity acquired by expanding the orbit to radius $2\lambda_L$ is 
Eq. (26), the
same as in a linear `thrust' over length $\lambda_L$ in direction perpendicular to the phase boundary \cite{copses}.

(ii) Within the theory of hole superconductivity, as discussed in numerous papers and for numerous reasons  \cite{holesc}, the
normal state carriers are necessarily holes.

The essential physics of the Meissner effect within the theory of hole superconductivity is orbit expansion driven by lowering
of quantum kinetic energy  \cite{kinetic1,kinetic2}. Instead, in conventional BCS theory, superconductivity is driven by lowering of potential energy.
We have argued  that no theory that explains superconductivity as driven by potential rather than kinetic energy can explain
the Meissner effect \cite{kinetic2}.

\section{ momentum in the electromagnetic field}
Within our theory, the superfluid charge density is slightly inhomogeneous, since the orbit expansion leads to higher negative charge density
within a London penetration depth of the surface, as shown schematically in Fig. 9. The excess charge density is given by \cite{electrospin}
\beq
\rho_{-}=en_s\frac{\hbar}{4m_e \lambda_L c}
\eeq
which gives rise to an outward pointing electric field in the interior of superconductors, that attains its maximum value $E_m$ near the surface, given by
\beq
E_m=-\frac{\hbar c}{4 e \lambda_L^2} .
\eeq
so that $E_m=-4\pi \rho_-\lambda_L$ \cite{electrospin}.
Therefore, there is electromagnetic momentum in the region within $\lambda_L$ of the surface where both electric and magnetic fields are present, according
to Eq. (24). The total electromagnetic angular momentum in that region, of volume $2\pi r_0 \lambda_L h$, is
\beq
\vec{L}_{em}=-\frac{1}{4\pi c} E_m H_c  r_0 (2\pi r_0 \lambda_L h)   \hat{z} .
\eeq
On the other hand, the extra mass density $\rho_-/e$ carries mechanical angular momentum, given by
(using Eqs. (57) and  (5))
\beq
\vec{L}_e^{extra}    =\frac{\hbar}{4m_e\lambda_Lc} \vec{L}_e=- \frac{\hbar}{4m_e\lambda_Lc}  \frac{m_e c}{2e}hr_0^2 H_c  \hat{z}
\eeq
so that \beq \vec{L}_{em}=-\vec{L}_e^{extra}\eeq
as required for momentum conservation.

It is interesting to note that for any value of $\rho_-$ Eq. (61) would hold, provided that $E_m=-4\pi \rho_-\lambda_L$ which is the condition
for the `surface charge density' $\sigma=\lambda_L\rho_-$ to screen the internal field $E_m$ so that it does not leak out of the superconductor: 
the angular momentum of the mass density $\rho_-/e$ moving at speed Eq. (26)  is exactly compensated by the angular momentum 
stored in the electromagnetic field.

In summary, the total electronic mechanical angular momentum of superconducting electrons in a magnetic field  is compensated by the angular momentum of the body plus 
a small contribution ($\sim 1/10^6)$ of  electromagnetic field momentum:
\beq
\vec{L}_e^{tot}=(1+\frac{\rho_-}{en_s})\vec{L}_e=-(\vec{L}_{body}+\vec{L}_{em})
\eeq
which completely accounts for momentum conservation and the mechanisms responsible for it.

             \begin{figure}
 \resizebox{6.5cm}{!}{\includegraphics[width=6cm]{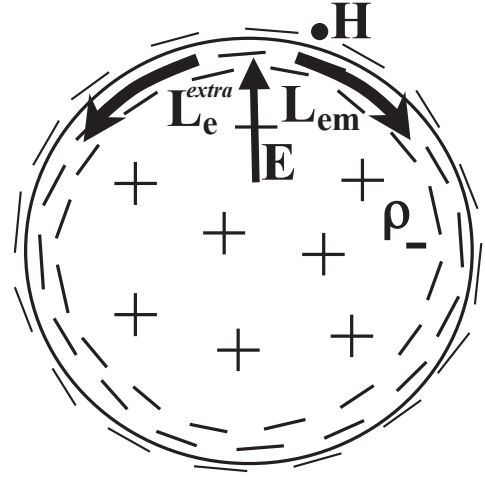}}
 \caption {Schematic view of superconducting region. Excess negative charge density $\rho_-$ resides within a London
 penetration depth of the phase boundary. A radial electric field exists in the interior. The extra 
 mechanical momentum
 $L_e^{extra}$  carried by
 the excess charge density $\rho_-$  is compensated by momentum $L_{em}$ in the electromagnetic field.
 }
 \label{figure1}
 \end{figure} 
 
\section{discussion}

In this paper we have argued that the only way that momentum can be transferred between the supercurrent and the body as a whole in a reversible way 
in processes where the normal-superconductor phase boundary moves is through mediation of the electromagnetic field,
which necessitates flow of charge in direction perpendicular to the phase boundary, and necessitates  hole carriers in the normal state.
Any alternative way to transfer the momentum between electrons and the body, i.e. scattering by impurities or phonons, would be an
irreversible process incompatible both with experiment and with established principles of superconductivity \cite{reversissue}.

We have furthermore argued that conventional BCS-London theory does not have the necessary physical elements to describe 
these processes in a reversible fashion. At the very least, it is a fact that no such description exists in the scientific literature. 

An important aspect of our explanation is that it only works if it is assumed that the mechanical momentum of an electron in the supercurrent is
\bmath
\beq
\vec{p}_e=-\frac{e\lambda_L}{c}H\hat{\theta}
\eeq
rather than 
\beq
\vec{p}_e=-\frac{m_e}{m*}\frac{e\lambda_L}{c}H\hat{\theta}
\eeq
\emath
as predicted by the conventional theory \cite{mstar}. The momentum formula Eq. (63b) works to explain momentum conservation in
process I, but cannot explain how momentum is conserved in processes II and III. The reason is, the explanation of
how momentum is transferred to the body  discussed in Sect. V involves momentum transfer
perpendicular to the motion, for which $m^*$ does not play a role: the momentum transferred to the ions Eq. (44) does not
depend on $m^*$. We have discussed elswhere \cite{mstar} other reasons for why the correct momentum expression
has to be Eq. (63a) rather than (63b).

BCS advocates argue that because at low temperatures  the superconducting state with the magnetic field excluded has lower free energy than the normal 
state with the magnetic field inside, the system will somehow `find its way' to the lower free energy state and expel the magnetic field.
They do not feel compelled to explain in the scientific literature $how$, within the confines of BCS theory
or even within time-dependent Ginzburg Landau theory, the process occurs respecting momentum conservation and reversibility.
We argue that such a stance is  unacceptable. One might say that it is equivalent to saying that
because an electron-positron pair has the same energy as a single 1.022 MeV photon, a theory predicting
that the former will decay into the latter has a claim to validity. It does not,   because such a process with a single photon
would violate momentum conservation, two photons are needed. Hence the theory with only one photon cannot be a valid
theory, no matter what other valid predictions it makes.

We argue that within BCS theory the Meissner  transition is a 
`forbidden' transition \cite{forbidden}, since the transition cannot take place respecting momentum conservation
if only the supercurrent changes its momentum. As in other forbidden transitions in physics, it is necessary to ascertain how long it will take the system
to get around the  selection rules originating in conservation laws by using higher order processes.
For example, consider gamma decay of excited atomic nuclei. Changes in the nuclear angular momentum by more than
one unit cannot occur by emission of a single  photon because this would violate angular momentum conservation. Changes by more than 1 unit can occur, but each additional unit of
spin change   inhibits the decay rate by about 5 orders of magnitude. For the highest known spin change of
8 units, the decay rate is suppressed by a factor $10^{35}$ and takes $10^{15}$ years  instead of $10^{-12}$ seconds.
Similarly we believe any route to explain the Meissner effect within BCS theory satisfying conservation laws and reversibility is highly `forbidden' and
would take time beyond the age of the universe for macroscopic systems. We would like to challenge BCS advocates  to show that
this is not so, by explaining the mechanism by which momentum is transferred between electrons and the body in a reversible fashion.

In contrast, the theory of hole superconductivity does have the physical elements necessary to explain these processes \cite{holesc}. In summary, those physical
elements, that are not part of BCS theory, are: (i) normal carriers are necessarily holes; (ii) when a system goes superconducting, not only the
occupation of Bloch states near the Fermi energy changes as predicted by BCS, keeping the individual Bloch states unaltered; instead, the
electronic wavefunction expands, and the highly dressed normal carrier becomes an undressed carrier with an extended wavefunction
that does not `see' the short-wavelength ionic potential \cite{holeelec2,sm}; (iii) as a consequence of (ii), supercarriers respond to external fields
according to the bare electron mass \cite{mstar} rather than the effective mass as predicted by BCS, and
(iv) as electrons become superconducting, negative charge extends beyond the normal-superconductor boundary
into the normal region.

Because these issues are basic and fundamental to the understanding of superconductivity, we argue that 
it is imperative to resolve them, and physicists should stop using conventional BCS-London theory to describe 
real superconductors unless or until it can be shown that the theory does not violate momentum conservation.

\appendix

\section{Derivation of the Hall coefficient and Hall force on the lattice}
We consider the Hall effect in the geometry of Fig. 7. The Hall coefficient is defined as
\beq
R_H=\frac{E_y}{J_x H}
\eeq
with $\vec{H}=H\hat{z}$ the applied magnetic field, $\vec{J}_x=J_x\hat{x}$ the current density and $\vec{E}_y=E_y\hat{y}$ the 
Hall field. The external force on an electron of wavevector $k$ in direction perpendicular to the current ($\hat{y}$ direction) is
\beq
F_{ext}^k=eE_y-\frac{e}{c}v_kH
\eeq
with
\beq
v_k=\frac{1}{\hbar}\frac{\partial \epsilon_k}{\partial k}
\eeq
We assume an isotropic band with energy $\epsilon_k$ and omit vector labels on the wavevectors. The total force on an electron of wavevector $k$ in direction
perpendicular to the current is
\beq
F_{tot}^k=    m_e\frac{d v_k}{dt}= \frac{m_e}{m^*_k}   \frac{d}{dt}(\hbar k)= \frac{m_e}{m^*_k}(eE_y-\frac{e}{c} v_k H)
\eeq
according to the semiclassical equation of motion, 
with
\beq
\frac{1}{m^*_k}=\frac{1}{\hbar^2}\frac{\partial ^2\epsilon_k}{dkdk}
\eeq
the effective mass tensor. On the other hand we can write the force on the electron of wavevector $k$ as the sum of the external force and the
force exerted by the lattice
\beq
F_{tot}^k=   F_{ext}^k + F_{latt}^k=  eE_y - \frac{e}{c}v_kH+ F_{latt}^k  .
\eeq
The total force on carriers per unit volume  in direction perpendicular to the current is, from integrating Eq. (A4) over the occupied states
\beq
F_{tot}\equiv \int_{occ}\frac{d^3k}{4\pi^3}F_{tot}^k=m_e e \int_{occ}\frac{d^3k}{4\pi^3}\frac{1}{m^*_k} (E_y-\frac{H}{c} v_k) 
\eeq
and the total force per unit volume exerted by the lattice on electrons in the transverse direction is, from Eq. A6)
\beq
F_{latt}\equiv \int_{occ}\frac{d^3k}{4\pi^3}F_{latt}^k  = F_{tot}-e \int_{occ}\frac{d^3k}{4\pi^3}  (E_y-\frac{H}{c} v_k)  .
\eeq
Next we evaluate Eqs. (A7) and (A8) for the cases of almost empty and almost full bands.

\subsection{Almost empty band}
The number of carriers and current are given by
\bmath
\beq
n_e= \int_{occ}\frac{d^3k}{4\pi^3}
\eeq
\beq
J_x=e \int_{occ}\frac{d^3k}{4\pi^3} v_k
\eeq
\emath
and we assume that for the occupied states near the bottom of the band
\beq
\frac{1}{m^*_k}\sim\frac{1}{m^*}>0
\eeq
independent of $k$. Eq, (A7) yields
\beq
F_{tot}=\frac{m_e }{m^*} (n_e e E_y-\frac{H}{c}J_x)
\eeq
and setting $F_{tot}=0$ yields
\beq
E_y=\frac{J_x H}{n_e e c}
\eeq
\beq
R_H=\frac{E_y}{J_x H}=\frac{1}{n_e e c}
\eeq
and from Eq. (A8)
\beq
F_{latt}=-(n_ee E_y-\frac{J_xH}{c})=0
\eeq
using Eq. (A12). Therefore, for this case, the Hall coefficient $R_H$ is negative,
 the total force exerted by the lattice on the carriers is zero, and conversely the total force
exerted by the carriers on the lattice is zero.

\subsection{Almost full band}
The number of carriers and current are given by
\bmath
\beq
n_h= \int_{unocc}\frac{d^3k}{4\pi^3}
\eeq
\beq
J_x= e\int_{occ}\frac{d^3k}{4\pi^3} v_k=-e \int_{unocc}\frac{d^3k}{4\pi^3} v_k
\eeq
\emath
and we assume
\beq
\frac{1}{m^*_k}\sim-\frac{1}{m^*}<0
\eeq
independent of $k$, for the unoccupied states near the top of the band. We have then
\beq
\int_{occ}\frac{d^3k}{4\pi^3}\frac{1}{m^*_k} = -\int_{unocc}\frac{d^3k}{4\pi^3}\frac{1}{m^*_k}=\frac{n_h}{m^*}
\eeq
\beq
e\int_{occ}\frac{d^3k}{4\pi^3}\frac{1}{m^*_k} v_k = -e\int_{unocc}\frac{d^3k}{4\pi^3}\frac{1}{m^*_k}v_k=   =-\frac{J_x}{m^*}
\eeq
hence from Eq. (A7)
\beq
F_{tot}=\frac{m_e }{m^*} (n_h e E_y+\frac{H}{c}J_x)
\eeq
and setting $F_{tot}=0$ yields
\beq
E_y=-\frac{J_x H}{n_h e c}
\eeq
\beq
R_H=\frac{E_y}{J_x H}=-\frac{1}{n_h e c}
\eeq
Therefore, in this case the Hall coefficient $R_H$ is positive. 
To find the force exerted by the lattice on electrons from Eq. (A8)  we use that
\beq
\int_{occ}\frac{d^3k}{4\pi^3}=\int_{zone}\frac{d^3k}{4\pi^3}-\int_{unocc}\frac{d^3k}{4\pi^3}=\frac{2}{v}-n_h
\eeq
with $v$ the volume of the unit cell, and use Eq. (A15b), and obtain
\beq
F_{latt}=-(eE_y(\frac{2}{v}-n_h)+\frac{H}{c}J_x)
\eeq
and using Eq. (A20)
\beq
F_{latt}=-\frac{2eE_y}{v}
\eeq
which unlike Eq. (A14) is $not$ zero. Hence, the total force per unit volume exerted by the carriers on the lattice is
\beq
F_{on-latt}= \frac{2eE_y}{v}
\eeq
Now the electric field $E_y$ also exerts a force on the lattice. The compensating ionic charge density per unit volume is
$|e|(2/v-n_h)$, hence the direct force of the electric field on the ions per unit volume is
\beq
F_{on-latt}^{E_y}=-eE_y(\frac{2}{v}-n_h)
\eeq
so that the $net$ force on the lattice per unit volume is
\beq
F_{on-latt}^{net}= F_{on-latt}+ F_{on-latt}^{E_y}=              n_h e E_y .
\eeq
or, using Eq. (A20)
\beq
\vec{F}_{on-latt}^{net}=-\frac{H}{c}J_x \hat{y} 
\eeq
in agreement with Eq. (31).

\section{The key issue of reversibility}
This paper rests on the   assumption that under ideal conditions the transition between normal and superconducting states in the presence
of a magnetic field is  
reversible, in other words, that it occurs  without change in the entropy of the universe. In this appendix we discuss the history of this issue,
and the experimental and theoretical evidence in its favor as it relates to our work.

Until the year 1933, when the Meissner effect was discovered \cite{meissner}, it was generally believed that  the transition from the superconducting
to the normal state when a current flows in the superconductor was necessarily irreversible: when the system became normal,
resistance would become  non-zero, the current would decay through the usual collision processes  that occur in the normal state, and  Joule heat $K$,
with $K$ the kinetic energy of the supercurrent, 
would be dissipated
in the process. As a consequence, the entropy of the universe would increase by an amount  $\Delta S_{irr}=K/T$, with $T$ the temperature.
No measurements were done to verify this assumption, presumably because it was considered to be a self-evident truth \cite{shoenberg}.  

The first hint that in fact this self-evident truth might $not$ be true  came from the experimental  finding \cite{rutgers}  of a relation between
the difference in specific heats in the normal and superconducting states and the temperature derivative of the 
critical magnetic field at the critical temperature:
\beq
C_s(T_c)-C_n(T_c)=\frac{V}{4\pi} T(\frac{\partial H_c(T)}{\partial T})^2_{T_c} 
\eeq
with $V$ the sample volume. This is known as the `Rutgers relation'. Already before the discovery of the Meissner effect, Gorter \cite{gorter} showed 
that Eq. (B1) follows from the assumption that the magnetic field $B$ is zero in the interior of superconductors $and$ that
the relation 
\beq
\frac {dQ}{T}=dS
\eeq
holds for the two phases, where $Q$ is the heat absorbed (released) in the transition  and $S$ the entropy of the phase. Eq. (B2)
is equivalent to saying that  the transition is reversible, hence that $no$ Joule heat is dissipated when the supercurrent stops.
Equation (B1) is simply derived from the relation  \cite{reif}
\beq
dF=-SdT-MdH
\eeq
for the free energies of the normal and superconducting phases assuming the magnetization $M=0$ for the normal phase and assuming
\beq
B=H+4\pi \frac{M}{V}=0 
\eeq
for  the superconducting  phase.  The above equations are  valid for a long cylinder with magnetic field in direction parallel to the axis. Using  
$S=-\partial F/    \partial T)_H$ it follows from (B3) and (B4)  that
\beq
S_n(T)-S_s(T)=\frac{L(T)}{T}=-\frac{V}{4\pi}H_c(T)\frac{\partial H_c(T)}{\partial T}
\eeq
along the phase transition line in the $H-T$ phase diagram. $L$ is the latent heat of the transition. Eq. (B5) is the analogous of the Clausius-Clapeyron equation relating pressure and temperature  for the liquid-solid or liquid-gas transition. It assumes that the free energies of the coexisting phases are
the same, $and$ that the transition is reversible. From $C=T\partial S/\partial T)_H$ and Eq. (B5)  it follows that
\beq
C_s -C_n  =\frac{V}{4\pi}T[(\frac{\partial H_c(T)}{\partial T})^2+ H_c(T)\frac{\partial^2H_c(T)}{\partial T^2}]
\eeq
along the coexistence curve. Eq. (B6) reduces to Eq. (B1) at the particular point $T=T_c$, where the magnetic field and the latent heat are zero.
All these relations follow from the fact that the relation between the difference in free energies in the normal and superconducting states at temperature  $T<T_c$ and the critical field $H_c(T)$ at that temperature is 
\beq
F_n(T)-F_s(T)=V\frac{H_c(T)^2}{8 \pi} .
\eeq
$if$  the transition is reversible \cite{gorter}. 

Note that Eq. (B7)   also follows from BCS theory \cite{tinkham}.
Therefore, within conventional BCS theory it is assumed, just as this paper assumes, that the normal-superconductor transition in a
magnetic field is a reversible phase transformation under ideal conditions.

These relations, together with London's electrodynamic equations, imply that the kinetic energy of the supercurrent is 
precisely given by the difference in the free energies of normal and superconducting states Eq. (B7). 
The supercurrent density  is given by
\beq
\vec{J}=en_s\vec{v}_s
\eeq
with $\vec{v}_s$  the superfluid velocity. London's equation is
\beq
\vec{\nabla}\times\vec{J}=-\frac{c}{4\pi \lambda_L^2}\vec{H}
\eeq
with $\lambda_L$ the London penetration depth. In a cylindrical geometry Eq. (B9) implies
\beq
J=-\frac{c}{4\pi \lambda_L}H
\eeq
so from Eqs. (B7) and (B10)
\beq
F_n-F_s=\frac{2\pi \lambda_L^2}{c^2}J^2
\eeq
and using the standard equation for the London penetration depth \cite{tinkham}
\beq
\frac{1}{\lambda_L^2}=\frac{4\pi n_s e^2}{m_e c^2}
\eeq
it follows that
\beq
F_n-F_s=\frac{n_s}{2}m_e v_s^2\equiv K
\eeq
The right-hand side of Eq. (B13) is the kinetic energy density of the supercurrent.
At the phase boundary between normal and superconducting phases
Eq. (B13) holds and this guarantees that there is phase equilibrium between the two phases \cite{londonh}.
Eq. (B13) also implies that when there is a small displacement of the phase boundary whereby a region goes from S to N, or from
N to S, the resulting change in the kinetic energy of the supercurrent is exactly compensated by the difference in the
free energies of the two phases. This implies that there is zero Joule heat dissipated when
the supercurrent stops.

After the Meissner effect was discovered  it would seem  very natural to expect that the transition in the presence of a magnetic field
was perfectly reversible for the following reason: if the kinetic energy of the supercurrent is stored rather than  dissipated as Joule heat
when the system becomes normal,  it will be
available to be converted again to kinetic energy of the supercurrent in the reverse transformation as the system becomes superconducting and
expels the magnetic field by generation
of the Meissner current.  
Nevertheless, despite the theoretical consistency and inherent beauty of the above considerations, it was thought necessary to check this
expectation experimentally, presumably because it was  considered  counterintuitive that a supercurrent could stop without any dissipation
when the system becomes normal.
 In the period 1934-1938 W. H. Keesom and coworkers did very extensive 
experimental work to check these predictions in a variety of ways \cite{keesom,keesom1,keesom2,keesom3,keesom4}. 
All the results found were consistent with the non-existence of   irreversible heat dissipation under ideal conditions
(e.g. pure samples,  the transition proceeding  slowly), and the thermodynamic relations
discussed above were found to hold to high accuracy.

Specifically, in Ref. \cite{keesom1}, Keesom and coworkers made calorimetric measurements along  a 
`Gorter cycle' \cite{gortercasimir,gorter} for $Tl$:  cooling below $T_c$ in zero field, then applying a field of magnitude just below the threshold value,
then heating across the transition in the presence of the field to above $T_c$, then switching off the field. 
They obtained the latent heat associated with the transition directly from the measurements, and applied the first and second laws of thermodynamics
to the heat exchanged along the cycle. Allowing for the possibility of an
irreversible increase in entropy $\sigma$ in the S-N transition, they  found $\sigma=0$ within experimental error, in other words that
\beq
\oint \frac{dQ}{T}=0
\eeq
holds in going around the cycle, hence concluded that  `{\it no irreversible entropy change occurs}'.  
In Ref. \cite{keesom2}, they repeated and confirmed these results to higher accuracy, stating that to be consistent with the experimental results 
`{\it it is essential that the persistent currents have been annihilated before the material gets resistance, so that no Joule heat
is developed}'.  In Ref. \cite{keesom3}, the atomic latent heat of $Sn$ was measured and compared with the theoretical expression
Eq. (B5) that assumes reversibility, finding that `{\it the agreement between the observed values and the calculated ones
is striking}'. In Ref. \cite{keesom4} both sides of Eq. (B6) were measured as well as of Eq. (B5) versus temperature, finding 
agreement `{\it under the assumption
that the transition from the superconductive to the non-superconductive
state is a reversible one}'. In Ref. \cite{keesom}, the reverse transition (N-S) in a magnetic field was also examined and it was found
that the latent heat in the N-S transition was equal to the one previously measured for the S-N transition, 
leading to  the conclusion that `{\it also in the
transition from the normal into the superconductive state no irreversible
increase in entropy takes place}. Quantitatively, the authors concluded from the measurements  that
`{\it the maximum limit of the irreversible increase in entropy comes to
$1.6\%$ at $3K$ and of $1\%$ at $2.6K$}' \cite{keesom}.

These experiments established that $at$ $most$  a  small fraction ($1-2\%$)  of the latent heat measured 
could be associated with irreversible processes.
Particularly at low temperatures, where the kinetic energy of the supercurrent becomes substantially larger 
than the latent heat, this  implies that not more than a tiny fraction ($<1\%$)  of the supercurrent could stop through onset of resistance with  dissipation of its
kinetic energy as Joule heat \cite{vanlaer}.

Several years later, Mapother \cite{map} again  tested the relations Eqs. (B5) and (B6) for $Sn$ and $In$.
He stated `{\it In this article we present the results of a careful comparison
between magnetic and calorimetric data for the
elements, Sn and In}.' `{\it It will be shown that the thermodynamic consistency between the two types of
measurement is, in general, of the order of $1\%$ and'
limited mainly by the precision of the calorimetric data.'} Thus, he 
established that  the relations Eqs. (B5), (B6)  hold to better than $1\%$, confirming reversibility.

These experiments and associated theory are extremely strong evidence  that the normal-superconductor transition in the presence of a magnetic
field   is reversible.  Note also that state of the art calorimetry \cite{calorim} is now substantially more advanced  than it was in the
1930's.
Thus, rather than to $1\%$ accuracy it may now be possible to establish experimentally 
that not more than $0.1\%$ or perhaps even not more than $0.01\%$ of the kinetic energy of the supercurrent  is dissipated  in irreversible processes.
This would imply that $99.99\%$ of the supercurrent stops without the current carriers undergoing
irreversible collisions. The question then would be, how is $99.99\%$ of the mechanical momentum of the supercurrent
transmitted to the body as a whole without irreversible collisions?  

We argue in this paper that to explain this  presents an insurmontable challenge to the conventional theory of superconductivity.
The conventional theory offers no mechanism by which {\it ``the persistent currents have been annihilated before the material gets
resistance, so that no Joule-heat is developed''}, as demanded by Keesom \cite{keesom2}.
The only way the conventional theory has addressed  this issue is by proposing that the momentum of the supercurrent 
is passed on to normal electrons when Cooper pairs dissociate,  that then transfer it to the body through collisions \cite{eilen}, which would have to be  perfectly elastic   and in addition generate no entropy.

So the conventional theory has to explain both how normal electrons can inherit the momentum of the supercurrent but not the
kinetic energy of the supercurrent (or at least not more than $0.01\%$ of the kinetic energy of the supercurrent), and  how
the momentum of these normal electrons can subsequently be transferred to the body as a whole  in a reversible way, without entropy generation. We believe it is impossible to do either without violating basic laws of physics, even if the transition occurs infinitely slowly.  
In any event, it certainly has not been done in the scientific literature to date.

\acknowledgements
The author is grateful to     D. J. Scalapino, A. J. Leggett, B. I. Halperin,  J. S. Langer and M. E. Fisher for discussions on these issues, 
and to   G. Sch\"on, N. D.  Goldenfeld,
V. Ambegaokar, E. Abrahams, L. Sham, N. Ashcroft, D. Mermin, D. Pines and P. W. Anderson   for comments.

\end{document}